\documentstyle[12pt]{article}
\def\be{\begin{equation}\label}
\def\ee{\end{equation}}
\def\bea{\begin{eqnarray}\label}
\def\eea{\end{eqnarray}}

\author{H.-J. Schmidt}

\title{The square of the Weyl tensor can be negative}
\date{}
\begin{document}
\maketitle

\begin{abstract}
\noindent 
We show that the square of the Weyl tensor can be negative by giving an  
 example: 
$$
ds^2 =  - dt^2   + 2yzdtdx  + dx^2 + dy^2 +  dz^2 \, . 
$$
This metric has the property that in a neighbourhood of  the origin, 
$$
C^{ijkl} \, C_{ijkl} < 0 \, . 
$$
\end{abstract}

\bigskip

\noindent 
KEY: Sign of curvature invariants; Weyl tensor 

\bigskip
\vspace*{2cm}
\bigskip
-------------------------

{\small 
\noindent H.-J. Schmidt, Institut f\"ur  Mathematik, Universit{\"a}t  
Potsdam, 

\noindent 
Am Neuen Palais 10, D-14469 Potsdam, Germany. 

\noindent  
E-mail: hjschmi@rz.uni-potsdam.de \quad
  http://www.physik.fu-berlin.de/\~{}hjschmi } 

\newpage 

Let us consider the metric
\be{y1}
ds^2 = - dt^2 \, + \, 2 \, y \,  z \, dt \, dx + dx^2 + dy^2 +  dz^2 \, .
\ee
At the coordinate origin, the metric equals  just the usual Minkowski
metric 
$$
\eta_{ij} = {\rm diag} (-1, \, 1, \, 1, \, 1)
$$
 and 
 all Christoffel symbols vanish. The only
components of $g_{ij,kl}$ which do not vanish are 
\be{y2}
g_{01,23} =g_{10,23} =g_{01,32} =g_{10,32}  =1 \, .
\ee
Therefore, at $t=x=y=z=0$, only those components of the Riemann 
tensor, where all 4 indices are different from each other, can have 
a non-vanishing  value.  Example: 
\be{y3}
R_{0312} = - 1/2\, ,  \qquad
  R^{0312} =  1/2  \, , \qquad  R_{0312} \cdot   R^{0312} = -  1/4\, .
\ee
One gets
$$
R^{ijkl} R_{ijkl} < 0 \, . 
$$
Further, the Ricci tensor vanishes there, and the Weyl tensor is
equal to the Riemann tensor. Consequently, the square of the Weyl tensor
is negative, and, by continuity, metric (\ref{y1}) represents an 
example of a spacetime, where 
$$
C^{ijkl} \, C_{ijkl} < 0 \,  , 
$$
  at least  in an open  neighbourhood of the origin.

\bigskip

Here  this inequality can  be applied:  Equation (27) of \cite{x1} reads
$$
B = \left( W^s_{\ ikl}\,  W^r_{\ abc} \, g_{sr} \, g^{ia}
 \, g^{kb} \, g^{lc} \right) ^{1/2}
$$
where $W^s_{\ ikl}$ denotes  the Weyl tensor, which we preferred to denote
by $C^s_{\ ikl}$  in the above text.  In the note added to \cite{x1}, 
the authors argued 
 that  eq. (27)  could not  have the character of a square root from 
a negative real.
 
The purpose of this comment is to clarify, that from the purely 
differential 
geometric point of view,  this is not the case, and therefore, 
one needs further
 physical motivations for the exclusion of negative values of the
 square of the Weyl tensor.

\bigskip

Note added: Possible negative  values  of  ``quadratic" curvature 
invariants 
have  also been discussed recently  in \cite{x2}; there the authors 
proposed to 
call  the sets  where this happens ``regions of gravitomagnetic 
dominance".

\end{document}